\documentclass[twocolumn,
 amsmath,amssymb,
 aps,
 prl
]{revtex4-1}

\usepackage{graphicx}
\usepackage{dcolumn}
\usepackage{bm}
\usepackage{braket}
\usepackage{dsfont}
\usepackage{color}

\begin{document}

\title{Superadiabatic Control of Quantum Operations}
\author{Jonathan Vandermause and Chandrasekhar Ramanathan}
\email{chandrasekhar.ramanathan@dartmouth.edu}
\affiliation{Department of Physics and Astronomy, Dartmouth College, Hanover, New Hampshire 03755, USA}
\date{\today}

\begin{abstract}
Adiabatic pulses are used extensively to enable robust control of quantum operations.  
We introduce a new approach to adiabatic control that uses the superadiabatic quality or $Q$-factor as a performance metric to design robust, high fidelity pulses. This approach permits the systematic design of quantum control schemes to maximize the adiabaticity of a unitary operation in a particular time interval given the available control resources.  The interplay between adiabaticity, fidelity and robustness of the resulting pulses is examined for the case of single-qubit inversion, and superadiabatic pulses are demonstrated to have  improved robustness to control errors. A numerical search strategy is developed to find a broader class of adiabatic operations, including multi-qubit adiabatic unitaries.  We illustrate the utility of this search strategy by designing control waveforms that adiabatically implement a two-qubit entangling gate for a model NMR system.  
\end{abstract}

\maketitle

\subsection{Introduction} 
\vspace*{-0.15in}

\noindent Speed and robustness are two essential characteristics of quantum control schemes that can often seem to be at odds with one another.  On one hand, fast diabatic gates designed using optimal control techniques can approach the quantum speed limit (QSL) and minimize errors due to decoherence \cite{Caneva-2009}.  However, these pulses are often sensitive to variations in the experimental control parameters and to uncertainties in the system Hamiltonian.  Additionally, the pulse shapes produced by these techniques are typically not smooth and frequently push the hardware limits of the system, requiring careful tuning and calibration to ensure high fidelity \cite{Borneman-2012,Martinis-2014}.  On the other hand, smoothly varying gates can be made robust to control errors and are typically easier to implement due to the simpler hardware requirements.  In particular, the utility of adiabatic gates that rely on the well-known adiabatic theorem \cite{Messiah} has been demonstrated for a variety of control tasks for quantum information processing \cite{Recati-2002,Duan-2003,Fuchs-2009,Chen-2014}.   
The transition-free driving of a quantum system enabled by adiabatic gates is particularly important if some 
excited states of the system are more susceptible to decoherence.  Hybrid approaches that combine both diabatic and adiabatic control have also been explored \cite{Bacon-2009,Hen-2015,Chasseur-2015}.

Amplitude- and frequency-modulated ``adiabatic pulses'' have long been used in nuclear magnetic resonance (NMR) to  efficiently invert nuclear spin states \cite{Silver,Pines,Garwood} and provide robustness against inhomogeneities in both the static and radiofrequency (RF) magnetic fields, finding applications in both high resolution NMR spectroscopy and \textit{in vivo} magnetic resonance imaging \cite{Tannus}.  Similar schemes have been used to optically control population transfers in atomic and molecular gases \cite{AllenEberly,Gaubatz-1990}.  

Finite time operations can only approximately satisfy the adiabatic condition, an issue that becomes  critical in the context of adiabatic quantum computation \cite{Farhi-2001}. 
Long control pulses are also susceptible to decoherence introduced by interactions with unwanted environmental degrees of freedom.  This raises an important question: what is the minimal time required to perform a high-fidelity adiabatic transition?  Counter-diabatic driving strategies --- called shortcuts to adiabaticity (STA) --- enable transition-less driving on much shorter timescales \cite{shortcuts}, even approaching the QSL \cite{Santos-2015}. 
These techniques have found applications in quantum state engineering \cite{Choi-2011,Choi-2012}, quantum computing \cite{Sarandy-2011,Herrera-2014}, many-body physics \cite{Saberi-2014} and quantum simulations \cite{delCampo-2011,delCampo-2012}, and have been shown to have robustness against control parameter variations \cite{Bason-2011}.  The DRAG pulses used in superconducting qubit implementations share many of these features as well \cite{Motzoi-2009,Gambetta-2011,Chow-2010}.
One challenge to implementing counter-diabatic driving strategies, particularly for systems of more than one-qubit, is that it may be difficult to generate the necessary counter-diabatic driving terms to ensure transition-free evolution using the available experimental controls.  

Here, we introduce a new approach to adiabatic control, based on Berry's ``superadiabatic'' formalism \cite{Berry}, that enables the systematic design of  quantum control schemes to maximize the adiabaticity of a unitary operation in a particular time interval, {\em given the available controls}.  We explicitly use the superadiabatic quality or $Q$-factor as a performance metric to optimize the available quantum control parameters.  The idea of a superadiabatic $Q$-factor was introduced by Deschamps et al.\  to explain the unexpectedly high fidelity of certain adiabatic pulses used in NMR \cite{Deschamps}.  We show that maximizing superadiabatic $Q$-factors improves the performance of standard one-qubit inversion pulses used in NMR and introduce a numerical search strategy to find a broader class of adiabatic unitaries when analytical solutions are not available. 
We numerically examine the interplay between adiabaticity, fidelity and robustness of the resulting pulses and show that superadiabatic pulses also improve robustness.  Finally, we show how the search technique can be used to create control waveforms that adiabatically implement a two-qubit entangling gate.  
While we explore these ideas in the context of NMR experiments, the ideas are broadly applicable to other modalities.

\vspace*{-0.15in}
\subsection{Superadiabatic $Q$-Factors} 
\vspace*{-0.15in}
\noindent Consider a time-dependent Hamiltonian $H_0(t)$ with instantaneous eigenbasis $\{{\ket{\lambda_0(t)}}\}$ at time $t$. 
Transforming to an interaction frame under the unitary operator $V_1 = \sum_\lambda{\ket{\lambda_0(t)}\bra{\lambda_0(0)}}$ that instantaneously diagonalizes the Hamiltonian yields an interaction frame Hamiltonian of the form $H_1 = D_1 + C_1$, where 
$D_1 = V_1^\dag H_0 V_1$ is diagonal and $C_1 = -i\hbar V_1^\dagger \dot{V_1}$ is a non-diagonal correction term (called an inertial term) arising from the time dependence of the Hamiltonian. 
A transition has typically been considered adiabatic if $||D_1(t)|| \gg ||C_1(t)||$ or $Q_1(t) \gg 1$ for the duration of the transition, where
\begin{equation}
Q_1(t) = \frac{||D_1(t)||}{||C_1(t)||}.
\label{eq:Q1}
\end{equation}
The ``adiabatic $Q$-factor" $Q_1$ is then defined as 
\begin{equation}
    Q_1 = \min\limits_{t \in [-\infty, \infty]}  Q_1(t).
\label{eq:Q1min}
\end{equation}
For finite-time processes, $C_1(t)$ is nonzero and the transformed Hamiltonian $H_1(t)$ is non-diagonal. In many STA approaches, a counter-diabatic driving term is introduced to explicitly cancel this non-diagonal inertial term \cite{shortcuts}.  Note that this is only possible if such an effective Hamiltonian can be generated with the available controls.

The above procedure for diagonalizing the instantaneous Hamiltonian can be applied to the transformed Hamiltonian $H_1$, yielding a new Hamiltonian $H_2$. Repeated indefinitely, this iterative procedure yields a countably infinite family of transformed Hamiltonians. Consider, for example, the Hamiltonian $H_{n-1}$. If the set $\{{\ket{\lambda_{n-1}(t)}}\}$ forms the instantaneous eigenbasis of $H_{n-1}$, the unitary operator $V_{n} = \sum_n{\ket{\lambda_{n-1}(t)}\bra{\lambda_{n-1}(0)}}$ diagonalizes $H_{n-1}$. In the interaction picture in which $H_{n-1}$ is instantaneously diagonalized, the Hamiltonian takes the form $H_{n} = D_{n} + C_{n}$, where $D_{n} = V_{n}^\dagger H_{n-1}(t) V_{n}$ and $C_{n} = -i\hbar V_{n}^\dagger \dot{V}_{n}$. By direct analogy with Eqs.\ (\ref{eq:Q1}) and (\ref{eq:Q1min}), the adiabatic $Q$-factor in frame $n$ takes the form
\begin{equation}
Q_{n} = \min\limits_{t \in [-\infty, \infty]} \frac{||D_{n}(t)||}{||C_{n}(t)||} \: \: \:.
\label{eq:Qi}
\end{equation}
Counter-diabatic driving STA strategies can also be derived for superadiabatic interaction frames \cite{Ibanez}.  

Deschamps {\em et al}.\ suggested that in a superadiabatic transformation, if the system starts out in one of the eigenstates of $H_n(0)$, it will evolve adiabatically to the target state in one of the superadiabatic frames as long as
\begin{equation}
Q_s \equiv \max\limits_{n \in \{1, 2, ...\}}Q_n \gg 1 \: \: \:,
\label{eq:Qs}
\end{equation}
where $Q_s$ is defined to be the \textit{superadiabatic $Q$-factor} \cite{Deschamps}.  

\vspace*{-0.15in}
\subsubsection{Scaling of $Q_1$}
\vspace*{-0.15in}

\noindent The $Q_1$ metric shows two important features:

\noindent 1.  If $H$ is a time-dependent Hamiltonian and $H'(t) = \alpha H(t)$, then $Q_1'(t) = \alpha Q_1(t)$ for $\alpha \in \mathbb{R}_+$.

\vspace*{0.1in}
\noindent Proof: Let $\{ \ket{n(t)} \}$ be the eigenvectors of $H(t)$. Then $\{ \ket{n(t)} \}$ are eigenvectors of $\alpha H(t)$, and hence 
\[V'(t) = \sum_{n} \ket{n(t)}\bra{n(0)} = V(t),\] from which we have that
\[ D'(t) = V'(t)H'(t)V'^{\dagger}(t) = \alpha D(t)\]
and
\[ C'(t) = -i \hbar V'^{\dagger}(t)\dot{V}'(t) = -i \hbar V^{\dagger}(t)\dot{V}(t) = C(t). \]
Therefore 
\[ Q_1'(t) = \frac{||D'(t)||}{||C'(t)||} =\frac{\alpha ||D(t)||}{||C(t)||} = \alpha Q_1(t).\]

\vspace*{0.1in}
\noindent 2.  If $H'(t) = H(\alpha t)$ where $t \in [0, \tau]$, then 
\[Q_1'(t) = Q_1(\alpha t)/\alpha\] for $\alpha \in \mathbb{R}_+$.

\vspace*{0.1in}
\noindent Proof: Let $u = \alpha t$. Then $V(u)$ is the unitary that diagonalizes $H'(t) = H(u)$, $D'(t) = D(u)$, and  
\[
\begin{split}
C'(t) = -i \hbar V'^{\dagger}(t) \dot{V'}(t) &= -i \hbar V^{\dagger}(u)\left(\frac{d}{dt}V(u)\right) \\ &= -i \hbar \alpha V^{\dagger}(u)\frac{d}{du}V(u) = \alpha C(u).
\end{split} \]
Therefore 
\[Q_1'(t) = \frac{||D'(t)||}{||C'(t)||} = \frac{||D(u)||}{\alpha ||C(u)||} = Q_1(\alpha t)/\alpha. \]

\vspace*{-0.15in}
\subsection{Analytical NMR Inversion Pulses} 
\vspace*{-0.15in}
\noindent  
To demonstrate the utility of the superadiabatic formalism, we examine the well-known \texttt{tanh/tan} adiabatic inversion pulse, one of a family of single spin-1/2 adiabatic inversion pulses used in NMR \cite{Hwang,Tannus}.  
For this system, the Hamiltonian during the pulse in a reference frame rotating at the nuclear spin Larmor frequency ($\omega_L$) takes the form:
\begin{equation}
H(t) =  \frac{\omega_1(t)}{2} \sigma_x + \frac{\Delta\omega(t)}{2} \sigma_z,
\label{eq:onespinHam}
\end{equation}
where $\Delta\omega = \dot{\phi}(t) - \omega_L$ is the resonance offset, $\phi(t)$ encodes the frequency and phase of the pulse, $\omega_1(t) = \gamma B_1(t)$, $\gamma$ is the nuclear gyromagnetic ratio, $B_1(t)$ is the amplitude of the applied RF field, and $\sigma_x$ and $\sigma_z$
are Pauli spin operators. Here and throughout this paper, $\hbar$ has been set to $1$. The goal of the pulse is to invert the state from $\ket{\uparrow} \equiv \ket{0}$ to $\ket{\downarrow} \equiv \ket{1}$.

 For a \texttt{tanh/tan} pulse of length $\tau$, the first half of the pulse ($t \le \tau/2$) can be described by  \cite{Hwang}:
\begin{equation}
\omega_1(t) = \omega_1^{\textrm{max}}\tanh{\left[ 2\xi t / \tau \right]}
\label{eq:w1}
\end{equation}
and
\begin{equation}
\Delta\omega(t) = A\frac{\tan\left[\kappa(1 - 2t/\tau)\right]}{\tan\kappa},
\label{eq:dw}
\end{equation}
where $\omega_1^{\textrm{max}}$ corresponds to the maximum RF field strength, and $\xi$, $\kappa$, and $A$ are parameters that can be optimized for a particular system.  For the second half of the pulse ($t > \tau/2$), $\omega_1(t) = \omega_1(\tau - t)$ and $\Delta\omega(t) = - \Delta\omega(\tau - t)$.

In the simulations here, the maximum RF amplitude was set at $\omega_1^{\textrm{max}} = 80$ krad/sec (12.7 kHz), a typical value for a liquid-state NMR spectrometer. This corresponds to a minimum gate time of 39.27 $\mu$s for a rectangular inversion pulse.  The remaining three parameters ($\xi$, $\kappa$, $A$) were numerically optimized using brute-force search to generate pulses that either (a) maximized the traditional adiabatic $Q$-factor $Q_1$, or (b) maximized the superadiabatic $Q$-factor $Q_s$. Since $s < 10$ for the pulse lengths examined, $Q_s$ was calculated by computing the maximum value of the first ten $Q$-factors, using the analytical forms derived for these pulses by Deschamps {\it et al} \cite{Deschamps}.  The optimal pulse parameters are shown in Table 1.

\begin{table}[h]
\small
	\begin{center}
    \begin{tabular}{ | l | c| c |c| }
    \hline
    pulse & $A$ (rad/sec) & $\kappa$ & $\xi$ \\ \hline
    $Q_1$ & $4.1\times 10^5$ & 6.9 & 16.1 \\ \hline
    $Q_s$ (120 $\mu$s) & $50.5\times 10^5$ & 65.8 & 49.2 \\ \hline
    $Q_s$ (50 $\mu$s) & $26.8\times 10^5$ & 36.3 & 41.6 \\ \hline
    \end{tabular}
\end{center}
\caption{Optimal pulse parameters for the \texttt{tanh/tan} pulse.}
\end{table}

\begin{figure*}
\includegraphics[scale = 0.61]{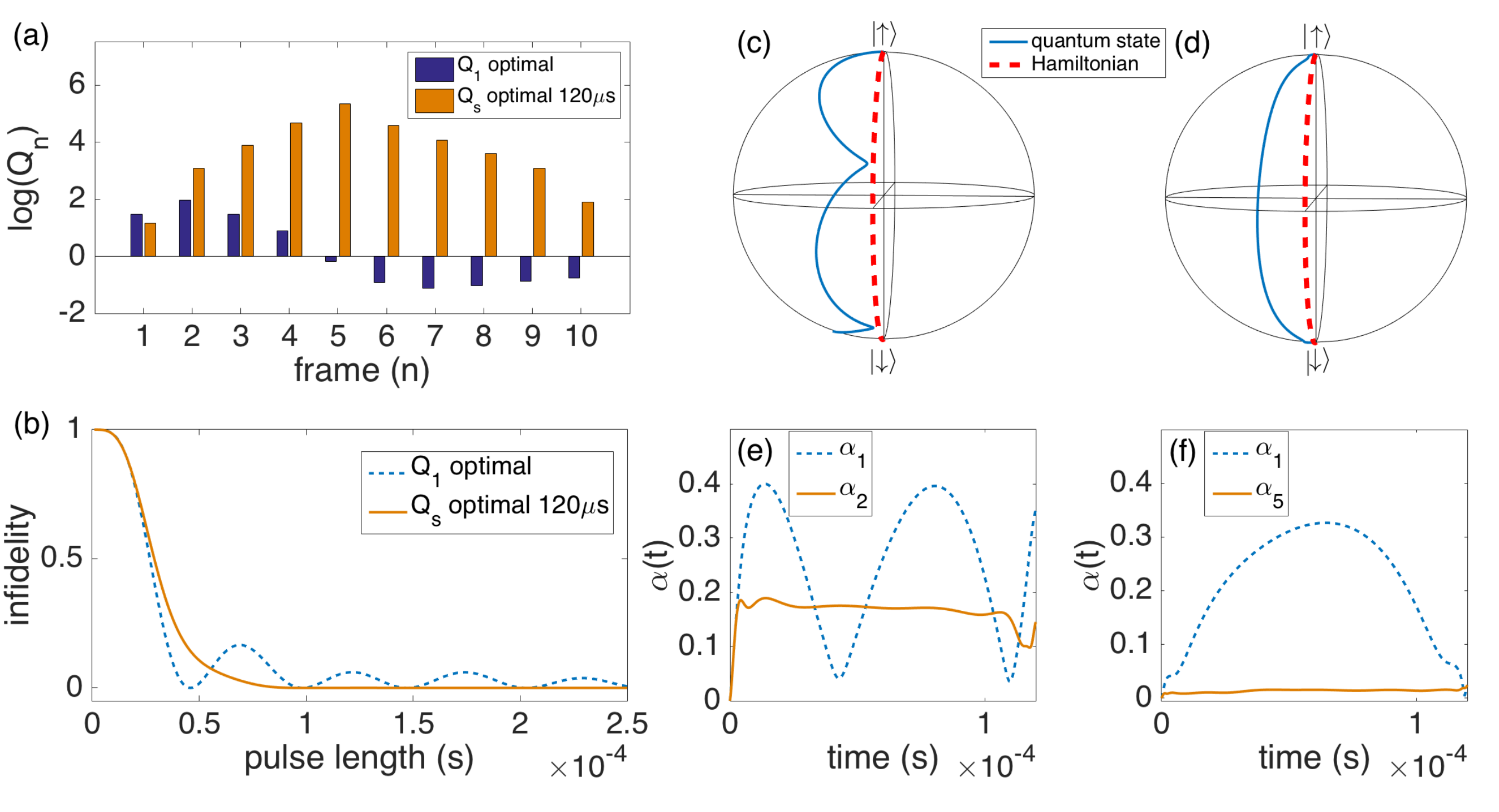}
\caption{A \texttt{tanh/tan} pulse of length $\tau = 120$ $\mu$s was optimized by varying $A$, $\kappa$ and $\xi$ in Eqs.\ (\ref{eq:w1}) and (\ref{eq:dw}) and setting $\omega_1^{\textrm{max}}$ to 80 krad/s. The resulting $Q_s$ optimized pulse is compared to the $Q_1$ optimal tanh/tan pulse.
In (a), the first ten adiabatic $Q$-factors, defined in Eq.\ (\ref{eq:Qi}), are plotted for both pulses on a log scale. 
We compare the performance of these two pulses by systematically reducing the pulse length, $\tau$. The infidelity of the inversion for each pulse length $\tau$ is plotted in (b). The quantum state's trajectory for the $Q_1$ optimized pulse is plotted in (c) and the trajectory of the $Q_s$ optimized pulse in (d).  The angles $\alpha_1(t)$ and $\alpha_s(t)$ are plotted as a function of time for the $Q_1$ ($s=2$) and the $Q_s$ ($s=5$) pulses in (e) and (f) respectively.}
\label{fig:fig1}
\end{figure*}

The optimization was first performed for pulse length $\tau = 120$ $\mu$s, about 3 times longer than the hard-pulse time. Figure \ref{fig:fig1}(a) compares the values of $\log{Q_n}$ (where $Q_n$ is defined by Eq.\ (\ref{eq:Qi})) for the two pulses at this pulse length. For both optimized pulses, $Q_n$ initially increases with $n$ until it reaches a peak value, which is the superadiabatic $Q$-factor, $Q_s$; in this case, for the $Q_1$ optimized pulse, $s = 2$, while for the $Q_s$ optimized pulse, $s = 5$. As Figure \ref{fig:fig1}(a) shows, for $n> s$, $Q_n$ begins to decrease or ``diverge,'' a phenomenon that has been attributed to the finite time of the transition \cite{Berry,Deschamps}.

The overall fidelity of the pulse was characterized by the overlap $F = |\braket{\psi(\tau)|1}|$.   Figure \ref{fig:fig1}(b) compares the performance, using the infidelity ($1-F^2$), of the two optimized \texttt{tanh/tan} pulse shapes as their duration was changed from 0 to 250 $\mu$s, demonstrating the improvement in fidelity provided by the superadiabatic pulse for pulse lengths $\tau > 56$ $\mu$s. Note the oscillations in the $Q_1$-optimized pulse that occasionally give very high fidelity at certain times.  

As a visual representation of the adiabatic dynamics, Figure \ref{fig:fig1}(c) and (d) show how the Bloch vector $\vec{v}(t)$ corresponding to the state tracks the Hamiltonian of the optimized pulses on the Bloch sphere for the $Q_1$ and the $Q_s$ optimized pulses respectively.  The time-dependent Hamiltonian can also be represented as a vector on the Bloch sphere $\vec{H}(t) = \omega_1(t) \hat{i} + \Delta\omega(t) \hat{k}$.  Since $\omega_1(t), \Delta\omega(t) \gg 1$ for most values of $t \in [0, \tau]$, we plot the projection of $\vec{H}(t)$ onto the Bloch sphere instead of $\vec{H}(t)$ itself.  The instantaneous deviation between $\vec{v} (t)$ and $\vec{H}(t)$ can be quantified in any superadiabatic frame by calculating the angle $\alpha_n(t)$ between $\vec{H}_n(t)$ and $\vec{v}_n(t)$ in that frame:
\begin{equation}
\alpha_n(t) = \arccos \left( \frac{\vec{H}_n(t) \cdot \vec{v}_n(t)}{||\vec{H}_n(t)|| \: ||\vec{v}_n(t)||} \right).
\label{eq:alphan}
\end{equation}
Figures \ref{fig:fig1}(e) and (f) show $\alpha_1(t)$ (dashed) and $\alpha_s(t)$ (solid) for the $Q_1$ optimized pulse ($s=2$) and the $Q_s$ optimized pulse ($s=5$) respectively. For the $Q_1$ optimized pulse, $\alpha_1$ and $\alpha_s$ are on the same order of magnitude, accounting for the quantum state's failure to reach the target state at this pulse length.  For the $Q_s$ optimized pulse, on the other hand, $\alpha_s(t)$ is negligible compared to $\alpha_1(t)$, suggesting that the state is locked to the superadiabatic Hamiltonian, $H_s$, but not to $H_1$. For $\tau$ = 120 $\mu$s, the infidelity of the $Q_1$ pulse is seen to be quite large, which is reflected in both Figures \ref{fig:fig1}(b) and (c).

We next examine the more general problem of engineering an optimally adiabatic pulse for a given pulse length $\tau$. As shown earlier $Q_1$ scales linearly with the length of the pulse if the pulse shape is held fixed, so a pulse shape that is $Q_1$-optimal for a particular pulse length $\tau$ will remain optimal for all pulse lengths. Importantly, this property does not hold for higher-order $Q$-factors, and hence a $Q_s$-optimal pulse at one pulse length $\tau$ may not be optimal at a different pulse length, suggesting that a separate optimization needs to be performed for each pulse length of interest. 

\begin{figure}
\includegraphics[scale = 0.4]{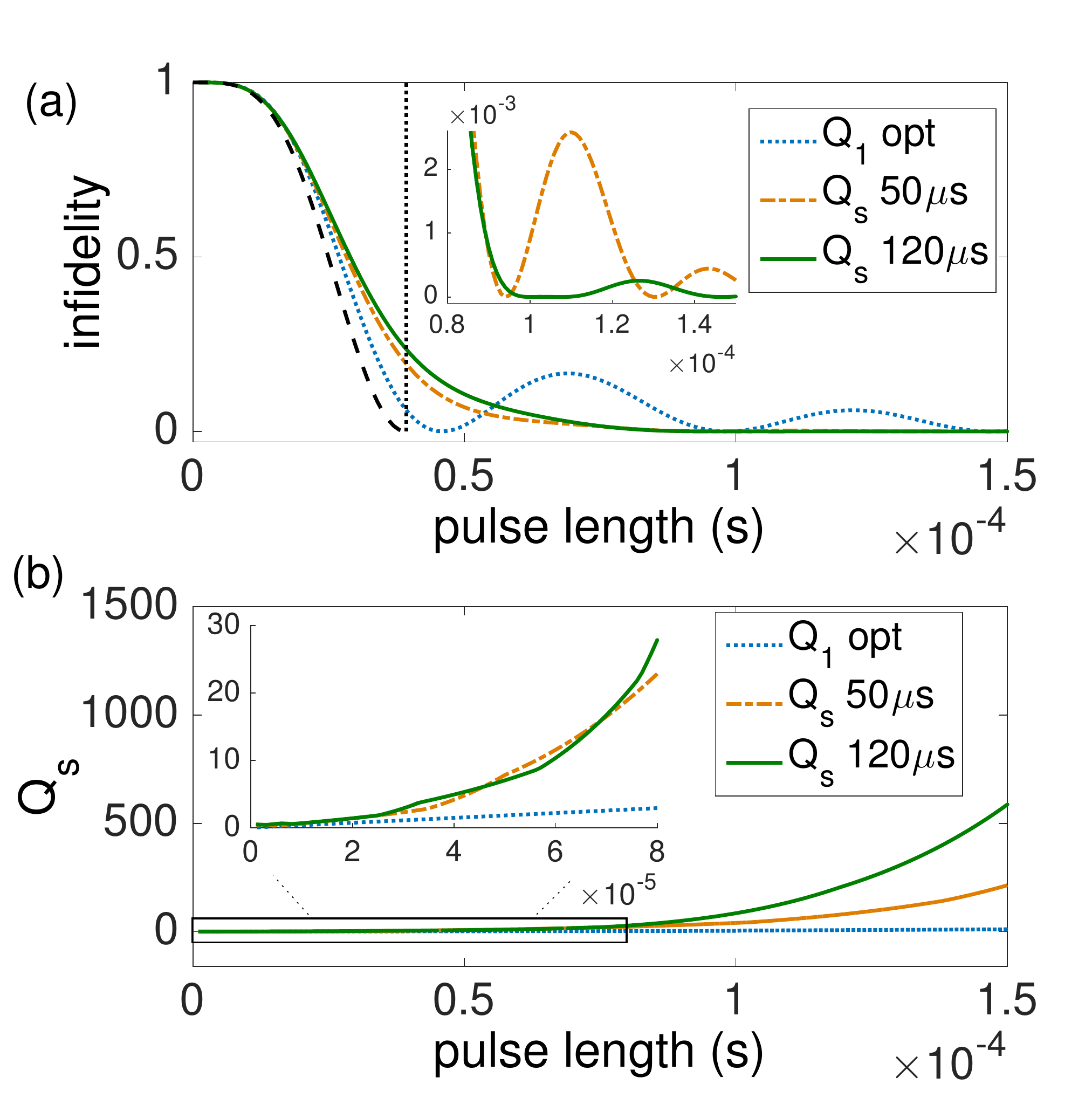}
\caption{(a) The infidelity of two $Q_s$-optimized pulses are compared with the $Q_1$ optimized pulse as a function of pulse length. The length of a hard $\pi$-pulse at $\omega_1 = 80$ krad/s is indicated by the vertical dotted line. The dashed black line plots the infidelity of the hard pulse as the pulse length is reduced to zero. The inset shows a magnified version of the plot in the range from 80 to 150 $\mu$s to show the improved performance of the pulse optimized for 120 $\mu$s at longer times. (b) $Q_s$ as a function of pulse length for each of the optimized pulses plotted in (a).  The inset shows the behavior of $Q$ as the pulse length approaches zero.}
\label{fig:fig2}
\end{figure}

Figures \ref{fig:fig2}(a) and (b) compare the performance of two $Q_s$ optimized \texttt{tanh/tan} inversion pulses that were optimized for inversion times of 50 $\mu$s and 120 $\mu$s to the original $Q_1$ optimized pulse.  The pulse designed for 50 $\mu$s is seen to perform better at shorter pulse lengths near 50 $\mu$s (in terms of both fidelity and superadiabatic $Q$-factor), while the pulse optimized for 120 $\mu$s performs better at longer times, with the behavior appearing to switch around 77 $\mu$s.  The dotted vertical vertical line in Figure \ref{fig:fig2}(a) indicates the duration of a ``hard'' rectangular $\pi$ pulse using the maximum available RF field of 80 krad/s, and the dashed line shows the fidelity achieved with this pulse.
The fidelities of the three adiabatic pulses approach that of the ideal hard pulse at short times, but never exceed it.  However the adiabaticity of the pulses is seen to rapidly fall as the pulse durations are reduced.  For these single qubit inversion pulses, we found that $Q_s \ge 10$  preserved the desired robustness properties for the adiabatic pulses.

\vspace*{-0.15in}
\subsection{Generalized Numerical Search Scheme} 
\vspace*{-0.15in}

\noindent  
 In the discussion above we considered the optimization of $Q_n$ for single-spin pulses of a specific analytical form.
In order to consider other unitaries, and to provide an optimization scheme that can be readily extended to higher-dimensional spaces where closed-form expressions for $Q_n$ are generally not available, we have designed an evolutionary search strategy that iterates on an initial guess pulse to produce numerically optimized pulse shapes that maximize $Q_n$ for any frame of interest $n$.  It should be noted that numerical optimization techniques have previously been used both to find the optimal pulse parameters of standard adiabatic NMR pulse shapes as well as to optimize arbitrary pulse shapes that maximize $Q_1$ \cite{Pines,Silver,Garwood}. The algorithm described below is similar to other derivative-free pulse-shaping methods that have been used in the past \cite{Warren}.

\begin{figure}
\includegraphics[scale = 0.4]{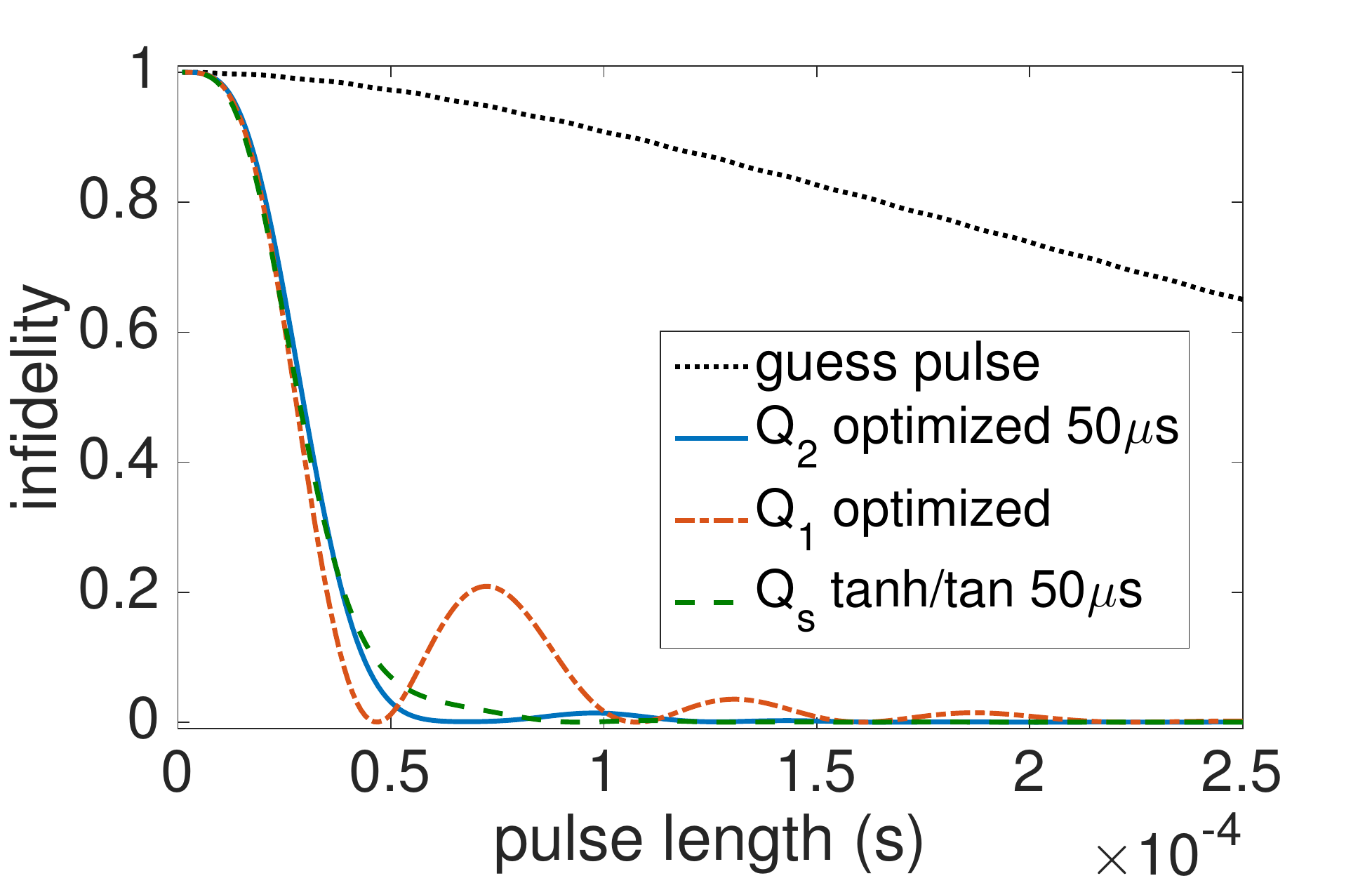}
\caption{The evolutionary strategy is applied to a guess pulse to optimize $Q_1$ and $Q_2$ sequentially. The infidelity of the resulting pulse, plotted as a solid line, shows improvement over the $Q_s$-optimal \texttt{tanh/tan} pulse of Figure \ref{fig:fig2}.}
\label{fig:fig3}
\end{figure}

We assume our Hamiltonian has the form $H(t) = H_0 + \sum_k u_k(t) H_k$, where $H_0$ is the time-independent part of the Hamiltonian and $u_k(t)$ are the control parameters corresponding to the Hermitian control operators $H_k$. Let $N$ be the number of time steps used to define the pulse.  An initial guess pulse (set of $u_k(t)$) is chosen that satisfies the necessary boundary conditions at $t=0$ and $t=\tau$ to ensure that the initial and final states are eigenstates of $H(0)$ and $H(\tau)$ respectively.  The key steps in our method are outlined here (see Appendix for additional details).
(i) The parameters of the guess pulse are perturbed in a time interval $[t_0 - \Delta,t_0+\Delta]$ and $Q_n$ re-calculated by numerically diagonalizing $H(t)$ to find all the $D_n$ and $C_n$ as outlined above.   Perturbations that improve $Q_n$ are preserved and used to update the guess pulse.  
(ii) The center of the perturbation ($t_0$), the size of the perturbed region ($2\Delta$) and the amplitude of the perturbation are all cycled systematically during the search as $Q_n$ is maximized. 

It is important to note that the evolutionary search does not guarantee convergence to a globally optimal pulse shape. As with many numerical search strategies, it is possible for the algorithm to get trapped in a local optimum. This may present a particular challenge as the size and complexity of the search space increases.


In Figure \ref{fig:fig3}, this search technique has been applied to the case of one-spin inversion. The chosen guess pulse consists of a linear ramp with arbitrarily chosen slope for the RF frequency offset $\Delta \omega(t)$ and a parabola for the RF amplitude $\omega_1(t)$ with zeros at the endpoints and a maximum value of $\omega_1^{\textrm{max}}$ at $t = \tau / 2$.
The evolutionary algorithm was first applied to the guess pulse to maximize $Q_1$. The fidelity profile of the resulting pulse is plotted as a dashed-dotted line in Figure \ref{fig:fig1}, showing considerable improvement over the guess pulse fidelity. This $Q_1$-optimized pulse was then used as the starting point for a second round of optimization, this time maximizing $Q_2$ at a pulse length 50 $\mu$s. The fidelity of the resulting pulse is also plotted in Figure \ref{fig:fig3} as a solid line. For comparison, the 50 $\mu$s $Q_s$-optimized \texttt{tanh/tan} pulse shown in Figure \ref{fig:fig2} is reproduced here as a dashed line. For pulse lengths around 50 $\mu$s, the numerically optimized pulse outperforms the $Q_s$-optimal \texttt{tanh/tan} pulse, demonstrating the potential benefits of numerical pulse-finding.

Optimal control techniques have also been used to maximize adiabaticity \cite{Rosenfeld-1996,Brif-2014}. 
Previous work using optimal control techniques to find adiabatic pulses used the integral of $Q_1$ as a global metric to search for pulses \cite{Rosenfeld-1996}.  It should be possible to adapt such a technique to maximize the integral of the superadiabatic $Q$-curve $Q_s(t)$ as well, which may enable the use of gradient based methods.  However, it is uncertain whether maximizing the integral of $Q(t)$ will preserve transition-free steering of the system at all times.

\vspace*{-0.15in}
\subsection{Robustness against inhomogeneity} 
\vspace*{-0.15in}

\noindent One of the principal benefits of adiabatic pulses is robustness against inhomogeneity in both the $\Delta\omega$ and $\omega_1$ terms of the Hamiltonian described by Eq.\ 4. Such robustness is desirable both for ensemble experiments in which there is a distribution of Hamiltonians (of both the system and control Hamiltonians either in space or in time), or if there is uncertainty in the Hamiltonian parameters.  We consider here the performance of the $Q_1$- and $Q_s$-optimized pulses discussed above when they are subjected to variations in both the frequency offset $\Delta\omega$ and the amplitude $\omega_1$. Consider a one-spin pulse described by the vector $\vec{\phi}(t) = \left[\Delta\omega(t), \omega_1(t)\right]$. We examine two distinct cases:
(i) The RF amplitude $\omega_1(t)$ is held fixed and a frequency offset term $\delta$ is added to $\Delta\omega(t)$, yielding the modified pulse \[ \vec{\phi}'(t) = \left[\Delta\omega(t) + \delta, \omega_1(t)\right]. \]
(ii) The original frequency offset $\Delta\omega(t)$ is preserved, but the RF amplitude $\omega_1(t)$ is multiplied by a scale factor $\sigma$, yielding the modified pulse \[ \vec{\phi}'(t) = \left[\Delta\omega(t), \sigma \omega_1(t)\right]. \]

\begin{figure}
\includegraphics[scale = 0.38]{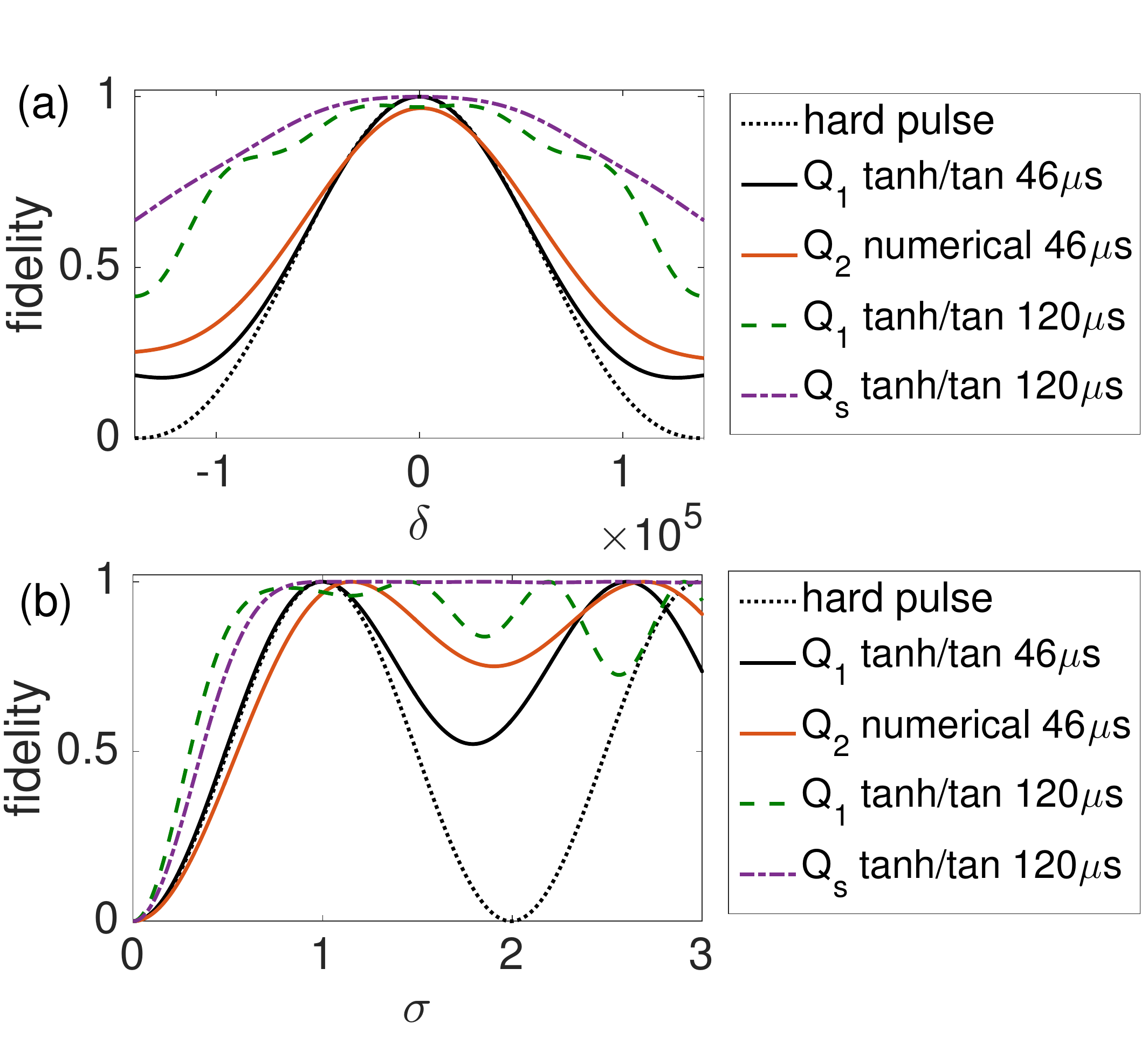}
\caption{Robustness of four optimized pulses to $B_0$ and $B_1$ inhomogeneity. Pulse performance is examined at two pulse lengths, 46 $\mu$s and 120 $\mu$s. The hard $\pi$-pulse fidelity is plotted as a dotted line. In (a), $\delta\omega(t)$ was subjected to a constant additive offset ranging from $-140$ krad/s to $140$ krad/s. In (b), $\omega_1(t)$ was subjected to a multiplicative offset ranging from 0 to 3. The pulses are more robust at the longer pulse length 120 $\mu$s. The superadiabatic pulses offer improved robustness with respect to both types of offsets.}
\label{fig:fig4}
\end{figure}

In Figure \ref{fig:fig4}, the robustness of several of the pulses discussed above is examined for offsets $|\delta| < 140$ krad/s and scalings $0 \le \sigma \le 3$. The dotted line corresponds to a hard $\pi-$pulse with an 80 krad/sec RF field and is the least robust of the pulses shown. The robustness of the $Q_1$ optimal \texttt{tanh/tan} pulse is examined at pulse length 46 $\mu$s, which is the shortest pulse length at which the pulse achieves nearly perfect fidelity (see Fig.\ 1(b) and 2(b)). The fidelity curves coincide for $|\delta| < 50$krad/s and $\sigma < 1.1$, suggesting that, for this pulse length, the $Q_1$-optimized pulse behaves like a hard pulse and confers little advantage in terms of robustness. The figure also compares the $Q_2$-optimized numerical pulse with the pulse length also set at 46 $\mu$s. Though the numerical pulse performs worse under ideal conditions, it achieves a higher fidelity than both the hard pulse and the $Q_1$-optimal pulse for $|\delta| > 36$ krad/s and $\sigma > 1.07$. The advantages of the adiabatic pulses are more pronounced for longer pulse lengths. The fidelities of the $Q_1$-optimized and 120 $\mu$s $Q_s$-optimized pulses at pulse length 120 $\mu$s are plotted as dashed lines. While both exhibit robustness for a wide range of offsets, the superadiabatic \texttt{tanh/tan} pulse outperforms the $Q_1$-optimized pulse for all $\delta$ and all $\sigma > 0.8$. Furthermore, the superadiabatic pulse achieves nearly perfect fidelity for $\sigma > 0.9$. This suggests that superadiabatic pulses offer an advantage not only in fidelity as a function of pulse length, as shown in Figures \ref{fig:fig1}--\ref{fig:fig3}, but also in robustness against variations in the Hamiltonian parameters.

\vspace*{-0.15in}
\subsection{Multiple qubits} 
\vspace*{-0.15in}
\noindent Our approach can be extended, in principle, to a larger number of qubits. However, since it requires the diagonalization of the instantaneous Hamiltonian to optimize the trajectory, it is not a scalable approach, a property it shares with most optimal control schemes. 
We consider a two-qubit system whose Hamiltonian is given by 
\begin{equation}
\begin{split}
H(t) = & \frac{\omega_1^{A}(t)}{2} \sigma_x \otimes \mathds{1} + \frac{\Delta \omega^{A}(t)}{2} \sigma_z \otimes \mathds{1} \\ 
+ & \frac{\omega_1^{B}(t)}{2} \mathds{1} \otimes \sigma_x + \frac{\Delta \omega^{B}(t)}{2} \mathds{1} \otimes \sigma_z \\
+ & \frac{\pi J}{2} \sigma_z \otimes \sigma_z,
\label{eq:twospinHam}
\end{split}
\end{equation}
where $\mathds{1}$ is the 2-by-2 identity operator, $\omega_1^{A,B}$ and $\Delta \omega^{A,B}$ are the qubit controls  for qubits $A$ and $B$ respectively, and $J$ is a fixed coupling constant in units of Hz.  This Hamiltonian arises in liquid-state NMR experiments and has also been implemented with superconducting qubits \cite{Plantenberg-2007}.  Here we demonstrate the use of our numerical strategy to adiabatically evolve a non-entangled pure state $\ket{\psi_i} =\ket{00}$ to the maximally entangled Bell state $\ket{\psi_t} = \frac{1}{\sqrt{2}} \left( \ket{00} + \ket{11} \right)$ without controlling $J$.  In some systems $J(t)$ can also be a time-dependent control when it can be experimentally varied \cite{Chen-2014}.

To design an adiabatic transition between $\ket{\psi_i}$ and $\ket{\psi_t}$, we must first identify an initial Hamiltonian $H(0)$ with eigenstate $\ket{\psi_i}$ and a final Hamiltonian $H(\tau)$ with eigenstate $\ket{\psi_t}$. Importantly, for the adiabatic theorem to hold, the two eigenstates must be non-degenerate and the order of the eigenstates must be preserved.  Setting $\omega_1^{A}(0) = \omega_1^{B}(0) = 0$ and requiring that $\Delta\omega^{A}(0) = \alpha > 0$ and $\Delta\omega^{B}(0) = -\beta < 0$, $H(0)$ can be written in matrix form as:
\[ \begin{split} 
 \left( \begin{array}{cccc}
\alpha - \beta + J & 0 & 0 & 0\\
0 & \alpha + \beta - J & 0 & 0\\
0 & 0 & -\alpha - \beta - J & 0\\
0 & 0 & 0 & -\alpha + \beta + J  \end{array} \right). \end{split} \]
If $\alpha > \beta$ and $\alpha, \beta > J$, the initial state $\ket{00}$ is the eigenvector of $H(0)$ corresponding to the second-largest eigenvalue $\alpha - \beta + J$.

The condition on $H(\tau)$ can be satisfied by setting $\Delta\omega^{A}(\tau) = \Delta\omega^{B}(\tau) = 0$ and further requiring that $\omega_1^{A}(\tau) = -A < 0$ and $\omega_1^{B}(\tau) = A > 0$. Again, $A$ is chosen so that $A > J$. In matrix form, with these conditions applied, $H(\tau)$ becomes:
\[ \begin{split} H(\tau) =
 \left( \begin{array}{cccc}
J & A & -A & 0\\
A & J & 0 & -A\\
-A & 0 & J & A\\
0 & -A & A & J  \end{array} \right), \end{split} \]
and the normalized eigenvector of $H(\tau)$ with the second-largest eigenvalue is
the Bell state $\ket{\psi_t} = \frac{1}{\sqrt{2}} \left( \ket{00} + \ket{11} \right)$.

Simulating a liquid state NMR experiment, we used a fixed value of 209.4 Hz for the $J$-coupling, corresponding to the measured proton-carbon coupling in a carbon-13 labeled chloroform sample.  For the initial guess pulse the RF amplitudes $\omega_1^{A}(t)$ (carbon) and $\omega_1^{B}(t)$ (proton) were chosen to vary linearly from $0$ krad/s at time $t = 0$ to $A = 78.5$ krad/s ($12.5$ kHz) at time $t = \tau$. The resonance offsets $\Delta\omega(t)^{A}$ and $\Delta\omega(t)^{B}$ were also chosen to be linear, with $\Delta\omega(\tau)^{A} = \Delta\omega(\tau)^{B} = 0$.
The values $\alpha = \Delta \omega^{A}(0) = 64$ krad/s and $\beta = -\Delta\omega^{B}(0) = 57$ krad/s were chosen to maximize $Q_1$ (See Appendix for details). The search algorithm was then used to iterate on this initial guess to find a control sequence that maximizes $Q_1$. The algorithm was carried out at an arbitrary pulse length since $Q_1$ scales linearly with the length of the pulse. 

\begin{figure}
\includegraphics[scale = 0.4]{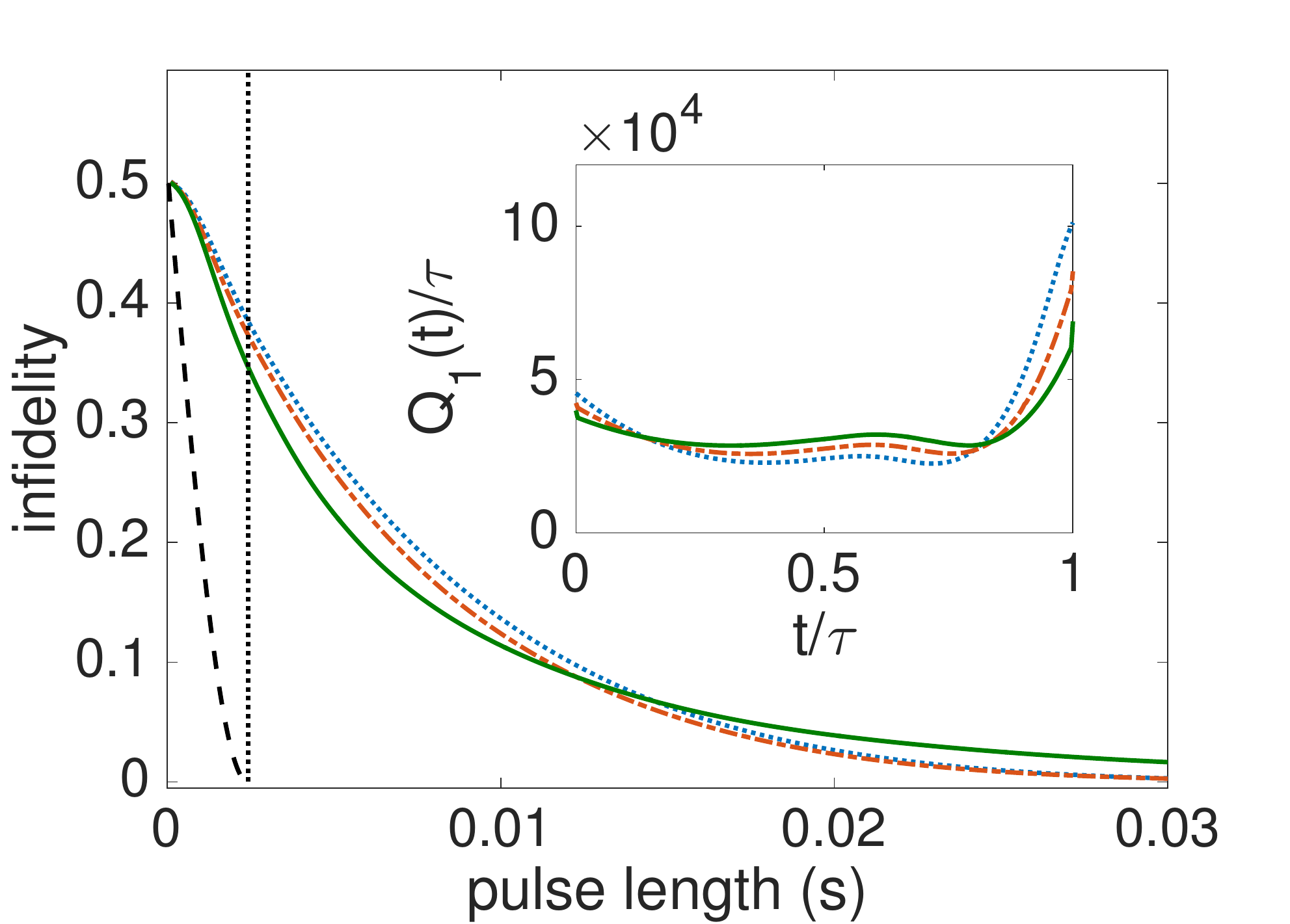}
\caption{Fidelity profiles of the two-qubit entangling pulse that takes the state $\ket{\psi_i} = \ket{00}$ to the state $\ket{\psi_t} = \frac{1}{\sqrt{2}}(\ket{00} + \ket{11})$. The evolutionary algorithm was applied to a linear guess pulse (dotted) for one round of optimization (dashed dotted) and three rounds (solid), where a round of optimization consists of each point in the pulse serving as the center of perturbation. Comparison is made to a diabatic gate that creates the same target state (dashed). The minimum time of the diabatic gate given $J = 209$ Hz is plotted as a vertical dotted line. Inset: $Q_1(t)$ is plotted for the three pulses, showing the algorithm's improvement in the first adiabatic $Q$-factor.}
\label{fig:fig5}
\end{figure}

The shape, performance, and fidelity of the resulting pulse depend on how long the algorithm is allowed to iterate on the initial guess pulse.  Here, a round of optimization is taken to be the number of times that each point in the pulse serves as a center of perturbation. 
Figure \ref{fig:fig5} shows the infidelity of the guess pulse and two $Q_1$-optimized pulses (following one and three rounds of optimization) as the length of the pulse is varied, while the inset plots $Q_1(t) = ||D_1(t)||/||C_1(t)||$ of both the guess pulse and optimized pulses, showing improvement in $Q_1 = \min{Q_1(t)}$. This improvement in adiabaticity is matched by an improvement in fidelity, with the optimized pulse outperforming the guess pulse for many of the depicted pulse lengths.  It is interesting to note that the QSL for a non-adiabatic gate in this two-qubit system is on the order of 1 ms with the same control resources (indicated by the dotted vertical line), which is significantly shorter than the high fidelity adiabatic pulses obtained here.  The non-adiabatic entangling gate consists of $\pi/2$ pulses on both spins, followed by a delay $1/2J$, which is then followed by a $\pi/2$ pulse on the protons.  The dashed line shows the drop in the fidelity of the non-adiabatic gate as the delay is reduced below $1/2J$.  It is the small size of the $J$-coupling that necessitates long adiabatic gates in this case.  

\begin{figure}
\includegraphics[scale = 0.35]{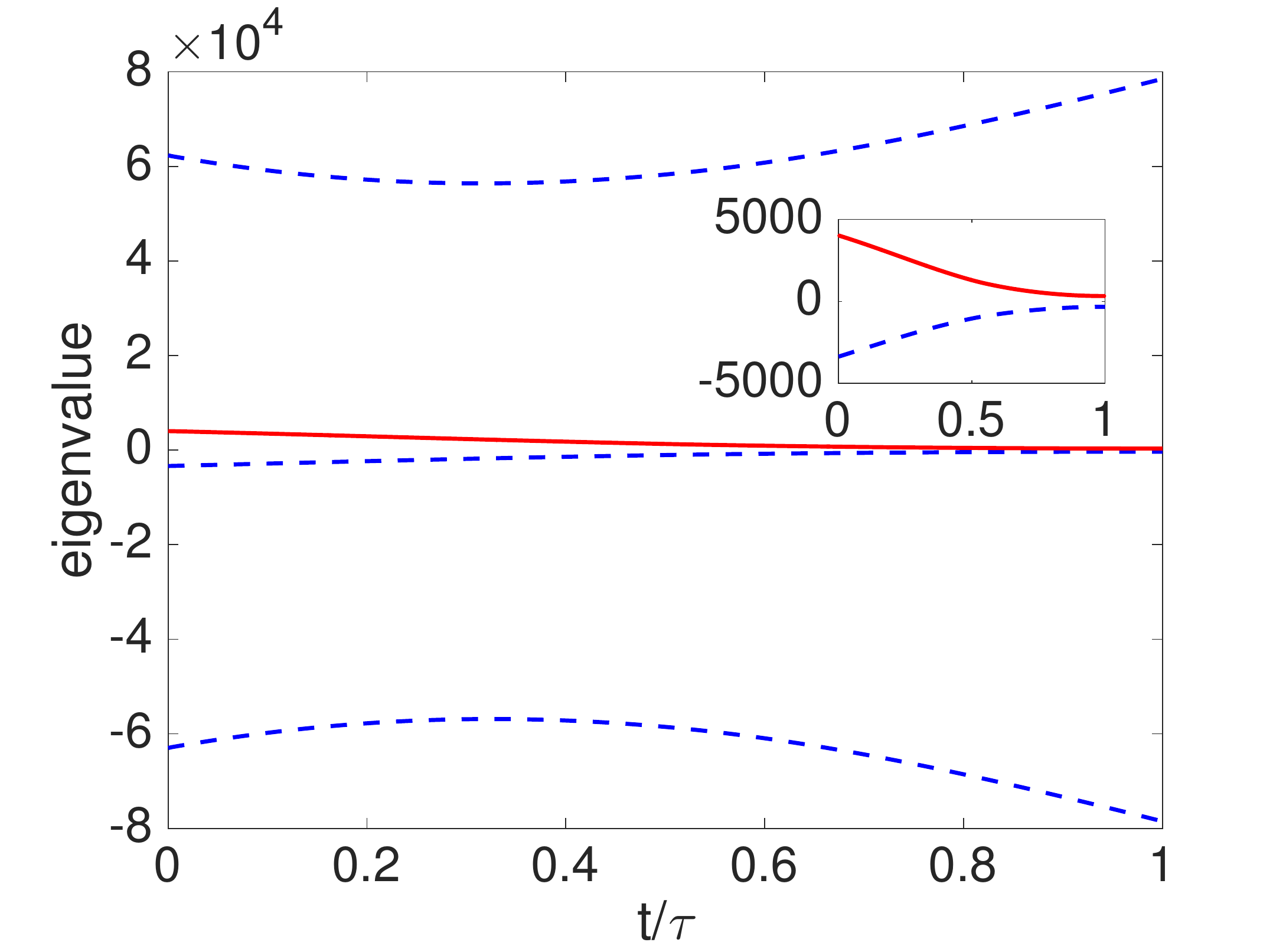}
\caption{Eigenvalues of the optimized two-qubit entangling Hamiltonian versus time. The adiabatic theorem requires that energy levels not cross, which is satisfied here. The eigenvalues corresponding to the trajectory between $\ket{00}$ at $t = 0$ and the Bell state at $t = \tau$ are plotted as a solid line. The eigenvalues are plotted in units of $\hbar / s$, using $\alpha = 64$ krad/s, $\beta = 57$ krad/s and $A = 78.5$ krad/s.}
\label{fig:fig6}
\end{figure}

Figure \ref{fig:fig6} shows how the instantaneous eigenvalues of the system change during the evolution of the final pulses (optimized 3 times) with the values of $\alpha,\beta$ and $A$ above.  The second largest eigenvalue, corresponding to the transition under consideration, is plotted as a solid line.  The figure confirms that the eigenvalues remain non-degenerate during the entire gate, with the size of the minimum energy gap set by the strength of the $J$-coupling.  We plan to explore superadiabatic control of multi-qubit systems in more detail in future work.

\subsection{Conclusions} 

\noindent We have introduced a new approach to transition-free driving of quantum systems.  This approach uses the superadiabatic $Q$-factor as a performance metric to design robust, high fidelity pulses that maximize the adiabaticity of the quantum operation in a particular interval, given the available experimental controls.  
These smoothly-varying super-adiabatic pulses are also easier to implement due to their typically simpler hardware requirements. 

For the case of single qubit inversion pulses, we found that optimizing $Q_s$ instead of $Q_1$ improved both  fidelity and robustness over a wide range of pulse lengths.   At shorter pulse lengths a trade-off was observed between fidelity and robustness, in which pulses that perform at high fidelity near the quantum speed limit tended to be less robust against inhomogeneity in the control parameters.  We also introduced a simple numerical search strategy to implement a broader class of adiabatic operations, including multi-qubit adiabatic unitaries, and designed an adiabatic control sequence to implement a two-qubit entangling gate.  
Our investigation highlights the generality of the $Q$-factor formalism, which can readily be extended to even larger Hilbert spaces or to systems characterized by entirely different Hamiltonians. 

The proposed method promises to offer another option in the toolbox of quantum control techniques.  Ultimately, it would be useful to systematically characterize the landscape of control in terms of potential trade-offs between desirable features such as speed, robustness against control errors and adiabaticity (or transition free driving).  This would allow experimentalists to tailor their control strategy to the specific experimental constraints in their setups.

\subsection{Acknowledgements} 
\noindent Jonathan Vandermause acknowledges support of a James O. Freedman Presidential Scholarship and a research grant from the Paul K. Richter and Evelyn E. Cook Richter Memorial Fund at Dartmouth College.  This research is based in part upon work supported by the National Science Foundation under CHE-1410504.

\bibliography{OptimizeAdiabatic_resub.bbl}

\begin{thebibliography}{19}%
\makeatletter
\providecommand \@ifxundefined [1]{%
 \@ifx{#1\undefined}
}%
\providecommand \@ifnum [1]{%
 \ifnum #1\expandafter \@firstoftwo
 \else \expandafter \@secondoftwo
 \fi
}%
\providecommand \@ifx [1]{%
 \ifx #1\expandafter \@firstoftwo
 \else \expandafter \@secondoftwo
 \fi
}%
\providecommand \natexlab [1]{#1}%
\providecommand \enquote  [1]{``#1''}%
\providecommand \bibnamefont  [1]{#1}%
\providecommand \bibfnamefont [1]{#1}%
\providecommand \citenamefont [1]{#1}%
\providecommand \href@noop [0]{\@secondoftwo}%
\providecommand \href [0]{\begingroup \@sanitize@url \@href}%
\providecommand \@href[1]{\@@startlink{#1}\@@href}%
\providecommand \@@href[1]{\endgroup#1\@@endlink}%
\providecommand \@sanitize@url [0]{\catcode `\\12\catcode `\$12\catcode
  `\&12\catcode `\#12\catcode `\^12\catcode `\_12\catcode `\%12\relax}%
\providecommand \@@startlink[1]{}%
\providecommand \@@endlink[0]{}%
\providecommand \url  [0]{\begingroup\@sanitize@url \@url }%
\providecommand \@url [1]{\endgroup\@href {#1}{\urlprefix }}%
\providecommand \urlprefix  [0]{URL }%
\providecommand \Eprint [0]{\href }%
\providecommand \doibase [0]{http://dx.doi.org/}%
\providecommand \selectlanguage [0]{\@gobble}%
\providecommand \bibinfo  [0]{\@secondoftwo}%
\providecommand \bibfield  [0]{\@secondoftwo}%
\providecommand \translation [1]{[#1]}%
\providecommand \BibitemOpen [0]{}%
\providecommand \bibitemStop [0]{}%
\providecommand \bibitemNoStop [0]{.\EOS\space}%
\providecommand \EOS [0]{\spacefactor3000\relax}%
\providecommand \BibitemShut  [1]{\csname bibitem#1\endcsname}%
\let\auto@bib@innerbib\@empty
\bibitem{Caneva-2009}{
T. Caneva, M. Murphy, T. Calarco, R. Fazio, S. Montangero, V. Giovannetti, G. E. Santoro, Phys. Rev. Lett., {\bf 103}, 240501 (2009).}
\bibitem{Borneman-2012}{
T. W. Borneman and D. G. Cory, J. Magn. Reson.,  {\bf 225}, 120--129 (2012).
}
\bibitem{Martinis-2014}{
J. M. Martinis and M. R. Geller, Phys. Rev. A, {\bf 90}, 022307 (2014).}
\bibitem [{\citenamefont {Messiah}(1965)}]{Messiah}%
  \BibitemOpen
  \bibfield  {author} {\bibinfo {author} {\bibfnamefont {A.}~\bibnamefont
  {Messiah}},\ }\href@noop {} {\emph {\bibinfo {title} {Quantum Mechanics}}}\
  (\bibinfo  {publisher} {North-Holland, Amsterdam},\ \bibinfo {year}
  {1965})\BibitemShut {NoStop}%
\bibitem{Recati-2002}{
A. Recati, T. Calarco, P. Zanardi, J. I. Cirac, and P. Zoller, 
Phys. Rev. A, {\bf 66}, 032309 (2002).
}
\bibitem{Duan-2003}{
L.-M. Duan and H. J. Kimble,
Phys. Rev. Lett., {\bf 90}, 253601 (2003).
}
\bibitem{Fuchs-2009}{
G.D. Fuchs, V.V. Dobrovitski, D.M. Toyli, F.J. Heremans, D.D. Awschalom,
Science, {\bf 326}, 1520--1522  (2009).}
\bibitem{Chen-2014}{
Y. Chen, C. Neill, P. Roushan, N. Leung, M. Fang, R. Barends, J. Kelly, B. Campbell, Z. Chen,
B. Chiaro, A. Dunsworth, E. Jeffrey, A. Megrant, J. Y. Mutus, P. J. J. O'Malley, C. M. Quintana, D. Sank,
A. Vainsencher, J. Wenner, T. C. White, M. R. Geller, A. N. Cleland, and J. M. Martinis, Phys. Rev. Lett., {\bf 113}, 220502 (2014).
}
\bibitem{Bacon-2009}{
D. Bacon and S.T. Flammia, Phys. Rev. Lett., {\bf 103}, 120504 (2009).}
\bibitem{Hen-2015}{
I. Hen, Phys. Rev. A, {\bf 91}, 022309 (2015).}
\bibitem{Chasseur-2015}{
T. Chasseur, L. S. Theis, Y. R. Sanders, D. J. Egger, F. K. Wilhelm, {Phys. Rev. A}, 
{\bf 91}, 043421 (2015).}
\bibitem [{\citenamefont {Silver}\ \emph {et~al.}(1984)\citenamefont {Silver},
  \citenamefont {Joseph},\ and\ \citenamefont {Hoult}}]{Silver}%
  \BibitemOpen
  \bibfield  {author} {\bibinfo {author} {\bibfnamefont {M.~S.}\ \bibnamefont
  {Silver}}, \bibinfo {author} {\bibfnamefont {R.~I.}\ \bibnamefont {Joseph}},
  \ and\ \bibinfo {author} {\bibfnamefont {D.~I.}\ \bibnamefont {Hoult}},\
  }\href@noop {} {\bibfield  {journal} {\bibinfo  {journal} {Journal of
  Magnetic Resonance}\ }\textbf {\bibinfo {volume} {59}},\ \bibinfo {pages}
  {347} (\bibinfo {year} {1984})}\BibitemShut {NoStop}%
\bibitem [{\citenamefont {Baum}\ \emph {et~al.}(1985)\citenamefont {Baum},
  \citenamefont {Tycko},\ and\ \citenamefont {Pines}}]{Pines}%
  \BibitemOpen
  \bibfield  {author} {\bibinfo {author} {\bibfnamefont {J.}~\bibnamefont
  {Baum}}, \bibinfo {author} {\bibfnamefont {R.}~\bibnamefont {Tycko}}, \ and\
  \bibinfo {author} {\bibfnamefont {A.}~\bibnamefont {Pines}},\ }\href@noop {}
  {\bibfield  {journal} {\bibinfo  {journal} {Phs. Rev. A}\ }\textbf {\bibinfo
  {volume} {32}},\ \bibinfo {pages} {6} (\bibinfo {year} {1985})}\BibitemShut
  {NoStop}%
\bibitem [{\citenamefont {Garwood}\ and\ \citenamefont
  {DelaBarre}(2001)}]{Garwood}%
  \BibitemOpen
  \bibfield  {author} {\bibinfo {author} {\bibfnamefont {M.}~\bibnamefont
  {Garwood}}\ and\ \bibinfo {author} {\bibfnamefont {L.}~\bibnamefont
  {DelaBarre}},\ }\href@noop {} {\bibfield  {journal} {\bibinfo  {journal}
  {Adv. Magn. Reson.}\ }\textbf {\bibinfo {volume} {153--177}},\ \bibinfo
  {pages} {155} (\bibinfo {year} {2001})}\BibitemShut {NoStop}%
\bibitem [{\citenamefont {Tann{\'u}s}\ and\ \citenamefont
  {Garwood}(1997)}]{Tannus}%
  \BibitemOpen
  \bibfield  {author} {\bibinfo {author} {\bibfnamefont {A.}~\bibnamefont
  {Tann{\'u}s}}\ and\ \bibinfo {author} {\bibfnamefont {M.}~\bibnamefont
  {Garwood}},\ }\href@noop {} {\bibfield  {journal} {\bibinfo  {journal} {NMR
  in Biomed.}\ }\textbf {\bibinfo {volume} {10}},\ \bibinfo {pages} {423}
  (\bibinfo {year} {1997})}\BibitemShut {NoStop}%
\bibitem [{\citenamefont {Allen}\ and\ \citenamefont
  {Eberly}(1987)}]{AllenEberly}%
  \BibitemOpen
  \bibfield  {author} {\bibinfo {author} {\bibfnamefont {L.}~\bibnamefont
  {Allen}}\ and\ \bibinfo {author} {\bibfnamefont {J.~H.}\ \bibnamefont
  {Eberly}},\ }\href@noop {} {\emph {\bibinfo {title} {Optical Resonance and
  Two-Level Atoms}}}\ (\bibinfo  {publisher} {Dover, New York},\ \bibinfo
  {year} {1987})\BibitemShut {NoStop}%
\bibitem [{\citenamefont {Gaubatz}\ \emph {et~al.}(1990)\citenamefont
  {Gaubatz}, \citenamefont {Rudecki}, \citenamefont {Schiemann},\ and\
  \citenamefont {Bergmann}}]{Gaubatz-1990}%
  \BibitemOpen
  \bibfield  {author} {\bibinfo {author} {\bibfnamefont {U.}~\bibnamefont
  {Gaubatz}}, \bibinfo {author} {\bibfnamefont {P.}~\bibnamefont {Rudecki}},
  \bibinfo {author} {\bibfnamefont {S.}~\bibnamefont {Schiemann}}, \ and\
  \bibinfo {author} {\bibfnamefont {K.}~\bibnamefont {Bergmann}},\ }\href
  {\doibase http://dx.doi.org/10.1063/1.458514} {\bibfield  {journal} {\bibinfo
   {journal} {J. Chem. Phys.}\ }\textbf {\bibinfo {volume} {92}},\ \bibinfo
  {pages} {5363} (\bibinfo {year} {1990})}\BibitemShut {NoStop}%
\bibitem{Farhi-2001}{
E. Farhi, J. Goldstone, S. Gutmann and M. Sipser, arXiv.org:quant-ph/0001106 (2001).}  
\bibitem [{\citenamefont {Torrontegui}\ \emph {et~al.}(2012)\citenamefont
  {Torrontegui}, \citenamefont {Ib{\'a}{\~n}ez}, \citenamefont
  {Mart{\'\i}nez-Garaot}, \citenamefont {Modugno}, \citenamefont {del Campo},
  \citenamefont {Gu{\'e}ry-Odelin}, \citenamefont {Ruschhaupt}, \citenamefont
  {Chen},\ and\ \citenamefont {Muga}}]{shortcuts}%
  \BibitemOpen
  \bibfield  {author} {\bibinfo {author} {\bibfnamefont {E.}~\bibnamefont
  {Torrontegui}}, \bibinfo {author} {\bibfnamefont {S.}~\bibnamefont
  {Ib{\'a}{\~n}ez}}, \bibinfo {author} {\bibfnamefont {S.}~\bibnamefont
  {Mart{\'\i}nez-Garaot}}, \bibinfo {author} {\bibfnamefont {M.}~\bibnamefont
  {Modugno}}, \bibinfo {author} {\bibfnamefont {A.}~\bibnamefont {del Campo}},
  \bibinfo {author} {\bibfnamefont {D.}~\bibnamefont {Gu{\'e}ry-Odelin}},
  \bibinfo {author} {\bibfnamefont {A.}~\bibnamefont {Ruschhaupt}}, \bibinfo
  {author} {\bibfnamefont {X.}~\bibnamefont {Chen}}, \ and\ \bibinfo {author}
  {\bibfnamefont {J.}~\bibnamefont {Muga}},\ }\href@noop {} 
  {\bibfield  {journal} {\bibinfo  {journal} {Advances In Atomic, Molecular, and Optical Physics}\
  }\textbf {\bibinfo {volume} {62}},\ \bibinfo {pages} {117--169} (\bibinfo
  {year} {2013})}  
\BibitemShut {NoStop}%
\bibitem{Santos-2015}{
A.C. Santos and M. S. Sarandy, Scientific Reports, {\bf 5}, 15775 (2015).}
\bibitem [{\citenamefont {Choi}\ \emph {et~al.}(2011)\citenamefont {Choi},
  \citenamefont {Onofrio},\ and\ \citenamefont {Sundaram}}]{Choi-2011}%
  \BibitemOpen
  \bibfield  {author} {\bibinfo {author} {\bibfnamefont {S.}~\bibnamefont
  {Choi}}, \bibinfo {author} {\bibfnamefont {R.}~\bibnamefont {Onofrio}}, \
  and\ \bibinfo {author} {\bibfnamefont {B.}~\bibnamefont {Sundaram}},\
  }\href@noop {} {\bibfield  {journal} {\bibinfo  {journal} {Phys. Rev. A}\
  }\textbf {\bibinfo {volume} {84}},\ \bibinfo {pages} {051601(R)} (\bibinfo
  {year} {2011})}\BibitemShut {NoStop}%
\bibitem [{\citenamefont {Choi}\ \emph {et~al.}(2012)\citenamefont {Choi},
  \citenamefont {Onofrio},\ and\ \citenamefont {Sundaram}}]{Choi-2012}%
  \BibitemOpen
  \bibfield  {author} {\bibinfo {author} {\bibfnamefont {S.}~\bibnamefont
  {Choi}}, \bibinfo {author} {\bibfnamefont {R.}~\bibnamefont {Onofrio}}, \
  and\ \bibinfo {author} {\bibfnamefont {B.}~\bibnamefont {Sundaram}},\
  }\href@noop {} {\bibfield  {journal} {\bibinfo  {journal} {Phys. Rev. A}\
  }\textbf {\bibinfo {volume} {86}},\ \bibinfo {pages} {043436} (\bibinfo
  {year} {2012})}\BibitemShut {NoStop}%
\bibitem [{\citenamefont {Sarandy}\ \emph {et~al.}(2011)\citenamefont
  {Sarandy}, \citenamefont {Duzzioni},\ and\ \citenamefont
  {Serra}}]{Sarandy-2011}%
  \BibitemOpen
  \bibfield  {author} {\bibinfo {author} {\bibfnamefont {M.~S.}\ \bibnamefont
  {Sarandy}}, \bibinfo {author} {\bibfnamefont {E.~I.}\ \bibnamefont
  {Duzzioni}}, \ and\ \bibinfo {author} {\bibfnamefont {R.~M.}\ \bibnamefont
  {Serra}},\ }\href@noop {} {\bibfield  {journal} {\bibinfo  {journal} {Phys.
  Lett. A}\ }\textbf {\bibinfo {volume} {375}},\ \bibinfo {pages} {3343}
  (\bibinfo {year} {2011})}\BibitemShut {NoStop}%
\bibitem [{\citenamefont {Herrera}\ \emph {et~al.}(2014)\citenamefont
  {Herrera}, \citenamefont {Sarandy}, \citenamefont {Duzzioni},\ and\
  \citenamefont {Serra}}]{Herrera-2014}%
  \BibitemOpen
  \bibfield  {author} {\bibinfo {author} {\bibfnamefont {M.}~\bibnamefont
  {Herrera}}, \bibinfo {author} {\bibfnamefont {M.~S.}\ \bibnamefont
  {Sarandy}}, \bibinfo {author} {\bibfnamefont {E.~I.}\ \bibnamefont
  {Duzzioni}}, \ and\ \bibinfo {author} {\bibfnamefont {R.~M.}\ \bibnamefont
  {Serra}},\ }\href@noop {} {\bibfield  {journal} {\bibinfo  {journal} {Phys.
  Rev. A}\ }\textbf {\bibinfo {volume} {89}},\ \bibinfo {pages} {022323}
  (\bibinfo {year} {2014})}\BibitemShut {NoStop}%
\bibitem [{\citenamefont {Saberi}\ \emph {et~al.}(2014)\citenamefont {Saberi},
  \citenamefont {Opatrn\'{y}}, \citenamefont {lmer},\ and\ \citenamefont {del
  Campo}}]{Saberi-2014}%
  \BibitemOpen
  \bibfield  {author} {\bibinfo {author} {\bibfnamefont {H.}~\bibnamefont
  {Saberi}}, \bibinfo {author} {\bibfnamefont {T.}~\bibnamefont {Opatrn\'{y}}},
  \bibinfo {author} {\bibfnamefont {K.~M.}\ \bibnamefont {lmer}}, \ and\
  \bibinfo {author} {\bibfnamefont {A.}~\bibnamefont {del Campo}},\ }\href@noop
  {} {\bibfield  {journal} {\bibinfo  {journal} {Phys. Rev. A}\ }\textbf
  {\bibinfo {volume} {90}},\ \bibinfo {pages} {060301(R)} (\bibinfo {year}
  {2014})}\BibitemShut {NoStop}%
\bibitem [{\citenamefont {del Campo}(2011)}]{delCampo-2011}%
  \BibitemOpen
  \bibfield  {author} {\bibinfo {author} {\bibfnamefont {A.}~\bibnamefont {del
  Campo}},\ }\href@noop {} {\bibfield  {journal} {\bibinfo  {journal} {Phys.
  Rev. A}\ }\textbf {\bibinfo {volume} {84}},\ \bibinfo {pages} {031606(R)}
  (\bibinfo {year} {2011})}\BibitemShut {NoStop}%
\bibitem [{\citenamefont {del Campo}\ \emph {et~al.}(2012)\citenamefont {del
  Campo}, \citenamefont {Rams},\ and\ \citenamefont {Zurek}}]{delCampo-2012}%
  \BibitemOpen
  \bibfield  {author} {\bibinfo {author} {\bibfnamefont {A.}~\bibnamefont {del
  Campo}}, \bibinfo {author} {\bibfnamefont {M.~M.}\ \bibnamefont {Rams}}, \
  and\ \bibinfo {author} {\bibfnamefont {W.~H.}\ \bibnamefont {Zurek}},\
  }\href@noop {} {\bibfield  {journal} {\bibinfo  {journal} {Phys. Rev. Lett.}\
  }\textbf {\bibinfo {volume} {109}},\ \bibinfo {pages} {115703} (\bibinfo
  {year} {2012})}\BibitemShut {NoStop}%
\bibitem{Bason-2011}{
M. G. Bason, M. Viteau, N. Malossi, P. Huillery, E. Arimondo, D. Ciampini, R. Fazio, V. Giovannetti, R Mannella, O. Morsch, Nature Physics, {\bf 8}, 147, (2012).}  
\bibitem{Motzoi-2009}{
F. Motzoi, J.M. Gambetta, P. Rebentrost, F.K. Wilhelm, Phys. Rev. Lett., {\bf 103}, 110501 (2009).}
\bibitem{Gambetta-2011}{
J.M. Gambetta, F. Motzoi, S.T. Merkel, F.K. Wilhelm, Phys. Rev. A, {\bf 83}, 012308, (2011).}
\bibitem{Chow-2010}{
J.M. Chow, L. DiCarlo, J.M. Gambetta, F. Motzoi, L. Frunzio, S.M. Girvin, R.J. Schoelkopf, Phys. Rev. A, {\bf 82}, 040305(R), (2010).} 
\bibitem [{\citenamefont {Berry}(1987)}]{Berry}%
  \BibitemOpen
  \bibfield  {author} {\bibinfo {author} {\bibfnamefont {M.}~\bibnamefont
  {Berry}},\ }\href@noop {} {\bibfield  {journal} {\bibinfo  {journal} {Proc.
  R. Soc. Lond. A}\ }\textbf {\bibinfo {volume} {414}},\ \bibinfo {pages} {31}
  (\bibinfo {year} {1987})}\BibitemShut {NoStop}%
\bibitem [{\citenamefont {Deschamps}\ \emph {et~al.}(2008)\citenamefont
  {Deschamps}, \citenamefont {Kervern}, \citenamefont {Massiot}, \citenamefont
  {Pintacuda}, \citenamefont {Emsley},\ and\ \citenamefont
  {Grandinetti}}]{Deschamps}%
  \BibitemOpen
  \bibfield  {author} {\bibinfo {author} {\bibfnamefont {M.}~\bibnamefont
  {Deschamps}}, \bibinfo {author} {\bibfnamefont {G.}~\bibnamefont {Kervern}},
  \bibinfo {author} {\bibfnamefont {D.}~\bibnamefont {Massiot}}, \bibinfo
  {author} {\bibfnamefont {G.}~\bibnamefont {Pintacuda}}, \bibinfo {author}
  {\bibfnamefont {L.}~\bibnamefont {Emsley}}, \ and\ \bibinfo {author}
  {\bibfnamefont {P.~J.}\ \bibnamefont {Grandinetti}},\ }\href@noop {}
  {\bibfield  {journal} {\bibinfo  {journal} {J. Chem. Phys.}\ }\textbf
  {\bibinfo {volume} {129}},\ \bibinfo {pages} {204110} (\bibinfo {year}
  {2008})}\BibitemShut {NoStop}%
\bibitem [{\citenamefont {Ib{\'a}{\~n}ez}\ \emph {et~al.}(2012)\citenamefont
  {Ib{\'a}{\~n}ez}, \citenamefont {Chen}, \citenamefont {Torrontegui},
  \citenamefont {Muga},\ and\ \citenamefont {Ruschhaupt}}]{Ibanez}%
  \BibitemOpen
  \bibfield  {author} {\bibinfo {author} {\bibfnamefont {S.}~\bibnamefont
  {Ib{\'a}{\~n}ez}}, \bibinfo {author} {\bibfnamefont {X.}~\bibnamefont
  {Chen}}, \bibinfo {author} {\bibfnamefont {E.}~\bibnamefont {Torrontegui}},
  \bibinfo {author} {\bibfnamefont {J.}~\bibnamefont {Muga}}, \ and\ \bibinfo
  {author} {\bibfnamefont {A.}~\bibnamefont {Ruschhaupt}},\ }\href@noop {}
  {\bibfield  {journal} {\bibinfo  {journal} {Phys. Rev. Lett.}\ }\textbf
  {\bibinfo {volume} {109}},\ \bibinfo {pages} {100403} (\bibinfo {year}
  {2012})}\BibitemShut {NoStop}%
\bibitem [{\citenamefont {Hwang}\ \emph {et~al.}(1998)\citenamefont {Hwang},
  \citenamefont {van Zijl},\ and\ \citenamefont {Garwood}}]{Hwang}%
  \BibitemOpen
  \bibfield  {author} {\bibinfo {author} {\bibfnamefont {T.-L.}\ \bibnamefont
  {Hwang}}, \bibinfo {author} {\bibfnamefont {P.~C.}\ \bibnamefont {van Zijl}},
  \ and\ \bibinfo {author} {\bibfnamefont {M.}~\bibnamefont {Garwood}},\
  }\href@noop {} {\bibfield  {journal} {\bibinfo  {journal} {J. Magn. Reson.}\
  }\textbf {\bibinfo {volume} {133}},\ \bibinfo {pages} {200} (\bibinfo {year}
  {1998})}\BibitemShut {NoStop}%
\bibitem{Warren}{
W.S. Warren and M.S. Silver,
Adv. Magn. Reson., {\bf 12}, 247, (1988).}
\bibitem{Rosenfeld-1996}{
D. Rosenfeld, Y. Zur, Magn. Reson. Med., {\bf 36}, 401--409 (1996).}
\bibitem{Brif-2014}{
C. Brif, M.D. Grace, M. Sarovar, K.C. Young, New J. Phys., {\bf 16}, 065013 (2014).}
\bibitem{Plantenberg-2007}{
J.H. Plantenberg, P.C. de Groot, C.J.P.M. Harmans and J.E. Mooij, Nature, {\bf 447}, 836 (2007).}


\end{thebibliography}%

\section{Appendix}

\subsection{Search Algorithm Details}
\begin{figure}
\includegraphics[scale = 0.32]{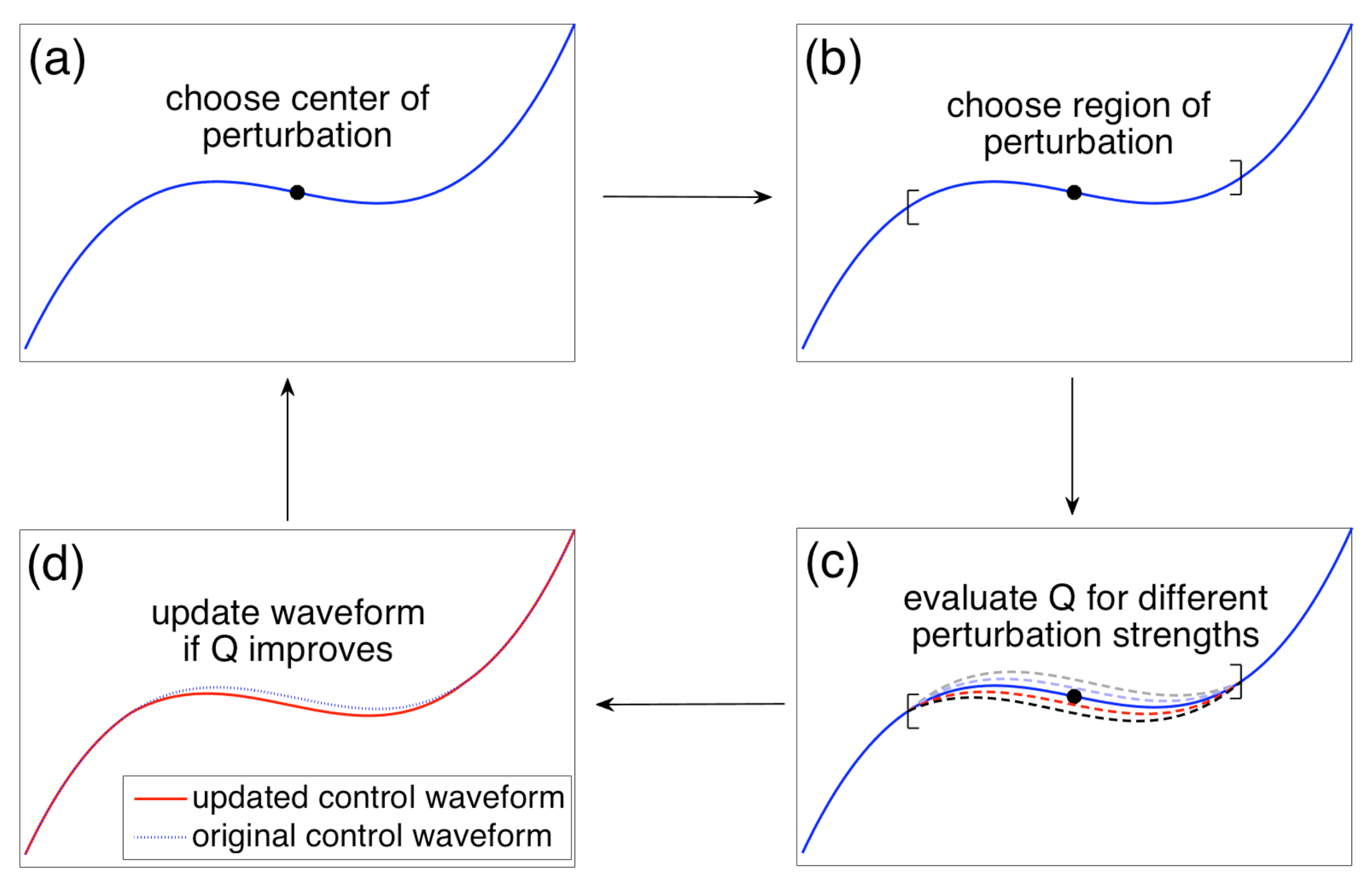}
\caption{The evolutionary strategy used to search for optimally adiabatic rf pulses, consisting of a four-step iterative procedure. (a) The center of perturbation is chosen. (b) The radius is chosen, defining an interval of perturbation. (c) The curve is perturbed and (d) the best perturbation is kept.}
\label{fig:fig7}
\end{figure}

\noindent Figure \ref{fig:fig7} outlines the steps of the algorithm.   Assume that the control pulses of length $\tau$ are divided into $N$ equal intervals $\Delta t$ such that $\tau = N\Delta t$.  The control waveform is parameterized by $u_k(t) = u_k(m \Delta t)$, $0 \le m \le N$.

\begin{enumerate}
\item[(a)] Choose the center of the perturbation ($m=\ell$) of the initial curve $u_k(m\Delta t)$ .

\item[(b)] Choose the radius of perturbation $r$.  For each center of perturbation $\ell$, the radii were allowed to vary from $r = N / 2$ (alters the entire curve) to $r = 2$ (smallest local perturbation).

\item[(c)] Introduce a \textit{parabolic perturbation} centered at $\ell$ with radius $r$: For every point $m \in \left[\ell - r, \ell + r \right]$, $u_k(m\Delta t)$ is changed to $\tilde{u}_k(m\Delta t)$ such that 
\[ \tilde{u}_k(m\Delta t) = u_k(m\Delta t) - \frac{\epsilon \left(m - (\ell - r)\right)\left(m - (\ell + r)\right)}{r^{2}}.  \]
where $\epsilon$ is a constant that controls the size of the perturbation.  For any given combination of $\ell$ and $r$, we perturb the curve ten times (chosen arbitrarily), in each case choosing $\epsilon$ to be a random value in the interval $[-\epsilon_{\textrm{max}}, \epsilon_{\textrm{max}}]$. The figure shows 4 such perturbations.

\item[(d)] The perturbation that maximizes the chosen adiabatic $Q$-factor is preserved. If, for any given radius of perturbation, none of the ten perturbations improved the adiabaticity, we return to step (b), this time choosing a smaller perturbation radius.

\end{enumerate}
When a perturbation that improves adiabaticity is found, the four-step procedure is repeated for a new center $\ell' = \ell + 1$ (mod N). If the algorithm does not find an improvement for any of the radii between $r = N/2$ and $r = 2$, the center is changed. Finally, since a pulse consists of two functions, $\Delta \omega(t)$ and $\omega_1(t)$, the algorithm toggled between the two: $\Delta \omega(t)$ was perturbed at center $\ell$, and before perturbing $\Delta \omega(t)$ again with $\ell' = \ell + 1$, $\omega_1(t)$ was perturbed at center $\ell$.

\subsection{Initial guess pulses for two-qubit control}

\begin{figure}[h]
\includegraphics[scale = 0.40]{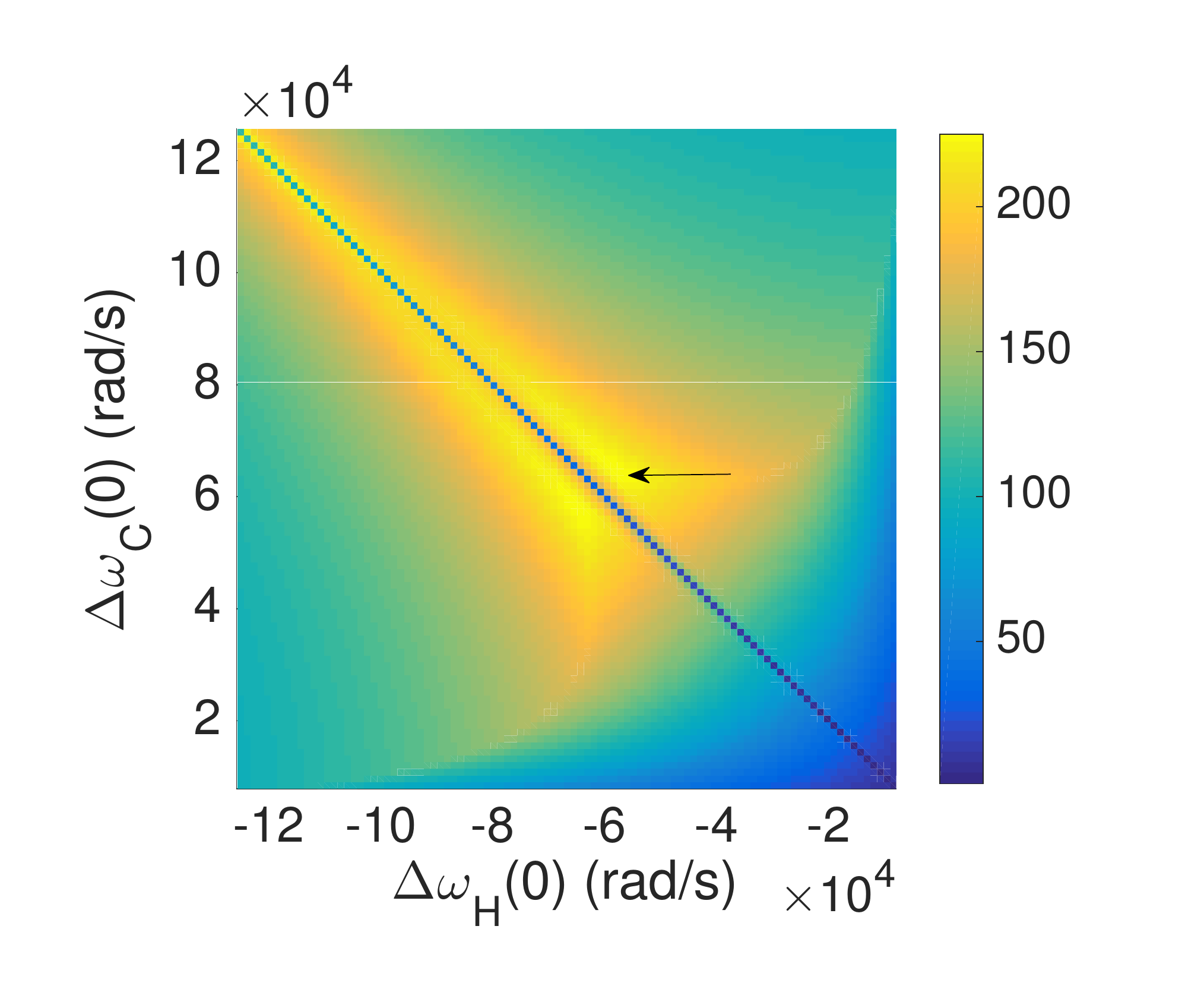}
\caption{$Q_1$ of guess pulse for different choices of $\Delta\omega^{C}(0)$ (carbon, system $A$) and $\Delta\omega^{H}(0)$ (hydrogen, system $B$). Lighter colors correspond to more adiabatic pulses. The black arrow indicates the guess pulse that was chosen.  The colorbar indicates the value of $Q_1$ for pulse length $10$ms.}
\label{fig:fig8}
\end{figure}

\begin{figure}[b]
\includegraphics[scale = 0.4]{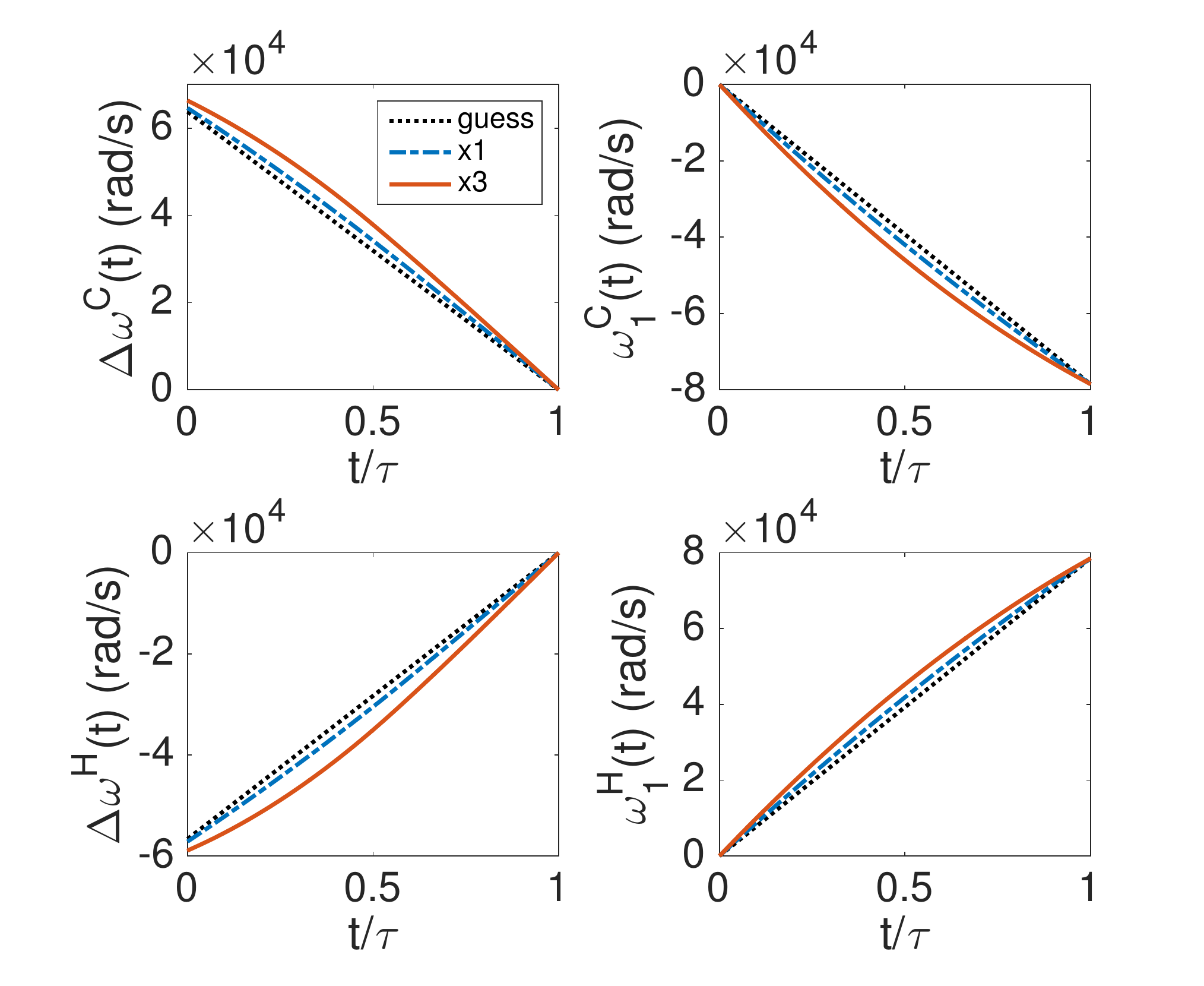}
\caption{The linear guess pulse (black dotted line) and two $Q_1$-optimized pulses, obtained by applying the evolutionary algorithm with each point as the center of perturbation once (blue dashed-dotted line) or three times (red solid line).}
\label{fig:fig9}
\end{figure}

\noindent Figure \ref{fig:fig8} motivates the choice of $\Delta\omega^{C}(0)$ (carbon, system $A$) and $\Delta\omega^{H}(0)$ (hydrogen, system $B$) mentioned in the text. Each point in the 2-D grid represents a different linear guess pulse.  The colorbar indicates the value of $Q_1$ -- the lighter the color of the grid point, the higher $Q_1$ is for the corresponding pulse.  The highest values of $Q_1$ occur when $\Delta\omega^{C}(0) = \Delta\omega^{H}(0)$, which is forbidden by the adiabatic theorem since it leads to degeneracy in the eigenvalues. Instead, the point indicated by the black arrow was chosen, with $\Delta\omega^{C}(0) \approx 64$ krad/s and $\Delta\omega^{H}(0) \approx -57$ krad/s. Figure \ref{fig:fig9} plots the shape of this linear guess pulse as well as the $Q_1$-optimized numerical pulse.

\end{document}